\documentclass[11pt, a4paper]{article}
\usepackage[a4paper,
    left=2cm,
    right=2cm,
    top=2.2cm,
    bottom=2.2cm
]{geometry}

\usepackage{setspace}
\usepackage{fancyhdr}
\usepackage[shortcuts]{extdash}
\usepackage{times}
\usepackage{booktabs,graphicx,siunitx,amsmath,amssymb}
\usepackage{caption}
\usepackage{subcaption}
\usepackage[utf8]{inputenc}
\usepackage{csquotes} 
\usepackage{hyperref}
\usepackage{authblk} 
\usepackage{palatino}
\usepackage[utf8]{inputenc}
\usepackage{hyperref} 

\usepackage[
  backend=biber,
  style=numeric,        
  citestyle=numeric-comp, 
  sorting=nyt
]{biblatex}
\title{Iran's Great Scientific Divergence: Counterfactual Evidence for the Long-Term Shock of the 1979 Revolution}

\date{\today}

\author[1]{Ehsan Roohi\thanks{Corresponding author: \href{mailto:roohie@umass.edu}{roohie@umass.edu}}}

\affil[1]{University of Massachusetts Amherst, Amherst, MA 01003, USA}

\begin{document}
\doublespacing
\maketitle

\begin{abstract}
This study quantifies the long-term impact of the 1979 Iranian Revolution on Iran's trajectory of scientific publications, a critical examination for science and public policy. Using comprehensive data from 1960 to 2024, we benchmark Iran against pre-revolutionary peers (e.g., South Korea) and employ multidimensional scientometrics, including the rigorous Synthetic Control Method (SCM). Results demonstrate a significant divergence: Iran, which led its peers in 1978, experienced collapse and stagnation (1980–1999) while peers grew exponentially. SCM, which optimally matches pre-1979 Iran with South Korea, quantifies a cumulative knowledge deficit of approximately 551,000 publications by 2024. Furthermore, despite a post-2000 volume recovery, a persistent quality gap exists; research impact (measured by Field-Weighted Citation Impact) consistently lags behind the global average. This analysis provides a robust, data-driven quantification of the generational opportunity cost of the 1979 disruption on national scientific development.
\end{abstract}

\textbf{Keywords:} Iranian Revolution; Cultural Revolution; Higher Education; Science Policy; Development; Knowledge Deficit; Counterfactual History; South Korea

\section{Introduction}

In the 1960s and 1970s, Pahlavi's Iran was defined by an ambitious, state-led modernization drive. Central to this national project, often articulated through development plans and concepts like the ``Great Civilization'' (Tammadon\-/e Bozorg), was the construction of a modern, Western-aligned scientific and higher education infrastructure. This was a deliberate national development strategy \autocite{vakilRevolutionaryEngineers2025}. It fueled the establishment and rapid expansion of elite institutions such as Pahlavi University in Shiraz, designed to be a national beacon, and the Aryamehr University of Technology (now Sharif), intended to train the engineering and technical cadre for a nascent industrial economy. This strategy was predicated on the belief that scientific capacity was synonymous with national sovereignty and long-term economic prosperity \autocite{King2004}.

The momentum of this pre-revolutionary push was not merely rhetorical. Bibliometric data from this era confirms a nascent system experiencing exponential growth. By the late 1970s, Iran was at a critical takeoff point. In a striking comparison that forms a central theme of this paper, Iran's total scholarly output had surpassed that of nations like South Korea and Taiwan \autocite{Amsden1989}. These countries, which shared a similar trajectory of state-led, Western-aligned development, would later become global exemplars of knowledge-based economies through targeted, long-term science and technology policies. The data from this period establishes a significant baseline: Iran's pre-1979 scientific trajectory was not only real but was on par with, or even leading, the very peers that would come to define the "Asian Tiger" model of development.

The 1979 Iranian Revolution, however, fundamentally altered this path. While all revolutions reshape national institutions, the impact on Iran's scientific enterprise was immediate, profound, and structural. The primary shock was the ``Cultural Revolution'' (Enqelab-e Farhangi), launched in 1980. With the stated goal of "Islamizing" universities and purging secular and Western influences, the policy resulted in the complete closure of all higher education institutions from 1980 to 1983. This three-year shutdown was not a pause; it was a systemic reset that severely reduced the nation's human capital. This event was the first and most direct cause of the collapse in Iran's scientific growth. As a direct result of the purges, ideological vetting, and forced exoduses that accompanied this period, the number of university teaching staff was nearly halved. According to data from the Ministry of Culture and Higher Education, the total number of teaching staff (including full-time, part-time, and sessional) fell from 16,877 in the 1979-80 academic year to 9,042 by 1982-83, creating a shortage of over 7,800 faculty members \autocite{Mojab1991}.

This internal, structural shock to the academic system was immediately exacerbated by a profound external crisis: the outbreak of the Iran-Iraq War (1980–1988). The war diverted all national resources and focus toward military defense and national survival. This paper aims to provide the first quantitative analysis of this combination of shocks—university closures, the loss of half the nation's professoriate, total war, and international isolation—which historical analysis suggests triggered a sharp decline in scientific productivity. We will empirically investigate the period from 1980 through the late 1990s, *termed here the "lost decade"*, to *test the hypothesis* that it constituted a period of deep stagnation and disconnection from the global scientific community.

Iran's subsequent journey—an initial collapse, a long stagnation, and then a rapid but incomplete recovery from the late 1990s onward\autocite{Stone2005},\autocite{hamdhaidari2008higher}\autocite{mozafari2016iran}\autocite{golkar2017politics}—does not fit neatly into the established models of scientific system recovery. Historical analysis reveals at least two distinct archetypes of recovery from systemic political shocks. The first can be described as a model of prolonged stagnation followed by slow, policy-driven reconstruction. The collapse of the Soviet Union provides a significant example; its formidable scientific enterprise endured a fifteen-year period of steady stagnation due to economic collapse and loss of institutional capacity. A recovery phase did not begin until 2006, initiated by new state policies that tied funding to research assessments \autocite{Yegorov2009}. The second archetype is one of a near-total reset followed by rapid, state-led exponential growth. China's trajectory after its Cultural Revolution (1966-1976) exemplifies this path. During this period, scientific work effectively came to a halt, with its output in top international journals dropping to nearly zero by 1973. Yet, immediately following the end of the revolution in 1976, its scientific output began a period of sustained, exponential growth, rebuilding from a near-zero baseline \autocite{Deng1997}. Iran's story, with its unique combination of a post-Soviet-style stagnation followed by a rapid, quantitative-focused resurgence, represents a distinct, hybrid pathway.

While historical narratives have captured the political dimensions of the revolution and the war, the long-term opportunity cost of this disruption on Iran's developmental trajectory remains largely unquantified. This paper situates Iran's interrupted journey within this comparative theoretical framework to better understand that cost. Using a scientometric approach, this paper moves beyond anecdote to provide a data-driven analysis of this historical divergence. It addresses two primary questions: 1) What was the immediate and long-term impact of the 1979 Revolution on Iran's scientific publication output compared to its peers? 2) What might Iran's scientific standing be today if its pre-revolutionary growth had continued uninterrupted?

We answer the second question by developing counterfactual models to quantify the "knowledge deficit." While a nation's scientific output is a robust indicator of its innovation potential and human capital \autocite{King2004}, this study is not merely a statistical exercise. By systematically analyzing the Iranian case against its peers, this paper provides a new, quantitative lens on the long-term developmental consequences of the 1979 Revolution. It seeks to explain how different initial conditions—including the nature of the political shock, the level of prior integration into global science, and the subsequent state ideology—lead to divergent pathways of scientific recovery, thereby refining our understanding of the compounding costs of historical disruption.

\section{Data and Methods}

\subsection{Data Source and Query}

The primary data source for this study is the Elsevier Scopus database, a comprehensive index of scholarly publications. We collected annual publication counts (limited to documents classified as ``articles'' or ``reviews'') for the period 1960–2024 for Iran and a carefully selected peer group. 

This comparative group was not chosen arbitrarily. It was designed to benchmark Iran's unique trajectory against diverse and relevant developmental models. The peer group includes:
\begin{itemize}
    \item {High-Growth ``Asian Tigers'' (South Korea, Taiwan, Singapore):} These nations serve as the most critical counterfactuals. Like pre-revolutionary Iran, they were state-led, Western-aligned ``developmental states'' that successfully executed long-term, strategic visions for building knowledge-based economies.
    \item {Stable Regional \& Mature Powers (Israel, Netherlands):} Israel provides a benchmark for a stable, high-performing regional scientific power, while the Netherlands represents a mature, established global scientific leader.
    \item {Other Comparative Cases (Greece, China):} Greece offers a European comparison with a different economic and integration model. China represents the archetype of a near-total systemic reset (after its Cultural Revolution) followed by explosive, state-led growth, providing a different recovery model.
\end{itemize}

The resulting annual publication counts form the basis of our longitudinal time-series analysis.

\subsection{Analytical Approach and Metrics}
Our study employs a multi-dimensional scientometric approach to provide a comprehensive assessment of Iran's scientific trajectory. The analysis integrates both quantitative and qualitative metrics to move beyond simple publication counts and capture a more nuanced picture of scientific development. The methodology is structured as follows:

\begin{enumerate}
    \item {Longitudinal Analysis of Publication Output:} We first analyze the growth trajectories of all eight countries in terms of publication volume across three distinct historical periods: 1960-1979 (Foundation), 1980-1999 (Divergence), and 2000-2024 (Recovery and New Order). This establishes the scale of the divergence.
    
    \item {Analysis of Scientific Impact and Quality:} To assess the quality and influence of the research, we incorporate two key normalized metrics from SciVal for the period 1996-2024:
    \begin{itemize}
        \item {Field-Weighted Citation Impact (FWCI):} To assess research quality and influence, we used the Field-Weighted Citation Impact (FWCI). This standard metric measures citation impact relative to the global average in the same field.
        \item {Outputs in Top 10\% Percentiles:} To measure the production of elite, highly-cited science, we analyzed the 'Outputs in Top 10\% Percentiles,' which represents the share of a country's papers that are among the top 10\% most cited globally.
    \end{itemize}
    This dual analysis of quantity and quality allows for a robust comparison of national scientific systems.
    
    \item {Counterfactual Modeling:} Finally, we develop and justify a suite of counterfactual models to quantify the "knowledge deficit" by projecting Iran's potential output under scenarios of uninterrupted growth.
\end{enumerate}

\subsection{Counterfactual Models}

The Counterfactual Models are described in Section~\ref{Counterfactual Analysis}. 

\subsection{Limitations of the Study}
This study has one primary limitation related to historical data coverage. Scopus's indexing of publications, particularly from the earlier decades of our analysis (1960s-1970s), may be less complete than for recent years. However, it is widely considered the most comprehensive database available for this type of longitudinal trend analysis, and any potential undercounting is likely to apply systemically across all countries in the dataset, thus preserving the validity of the comparative trends observed. The study's focus on both publication volume and normalized impact metrics provides a robust and multifaceted view of the scientific trajectories analyzed.
A second limitation pertains to the assumptions underlying the counterfactual models, particularly Model D (South Korea Growth-Proxy). While this model provides a robust, data-driven estimate of the ``knowledge deficit,'' it represents an idealized \textit{potential} trajectory. The realization of this trajectory assumes not only the absence of the post-1979 political disruption but also the sustained implementation of effective, long-term science and industrial policies analogous to those successfully adopted by South Korea's developmental state. The model quantifies the opportunity cost of the disruption, contingent upon optimal policy choices in the counterfactual timeline.

\section{Results: A Tale of Two Trajectories – Quantity and Quality}

Fig. 1 shows the number of publications of Iran during the period 1969–2024. Figs. 2–4 show the same statistics for South Korea, Taiwan, and Turkey.

\begin{figure}[htbp]
  \centering
  \includegraphics[width=\textwidth]{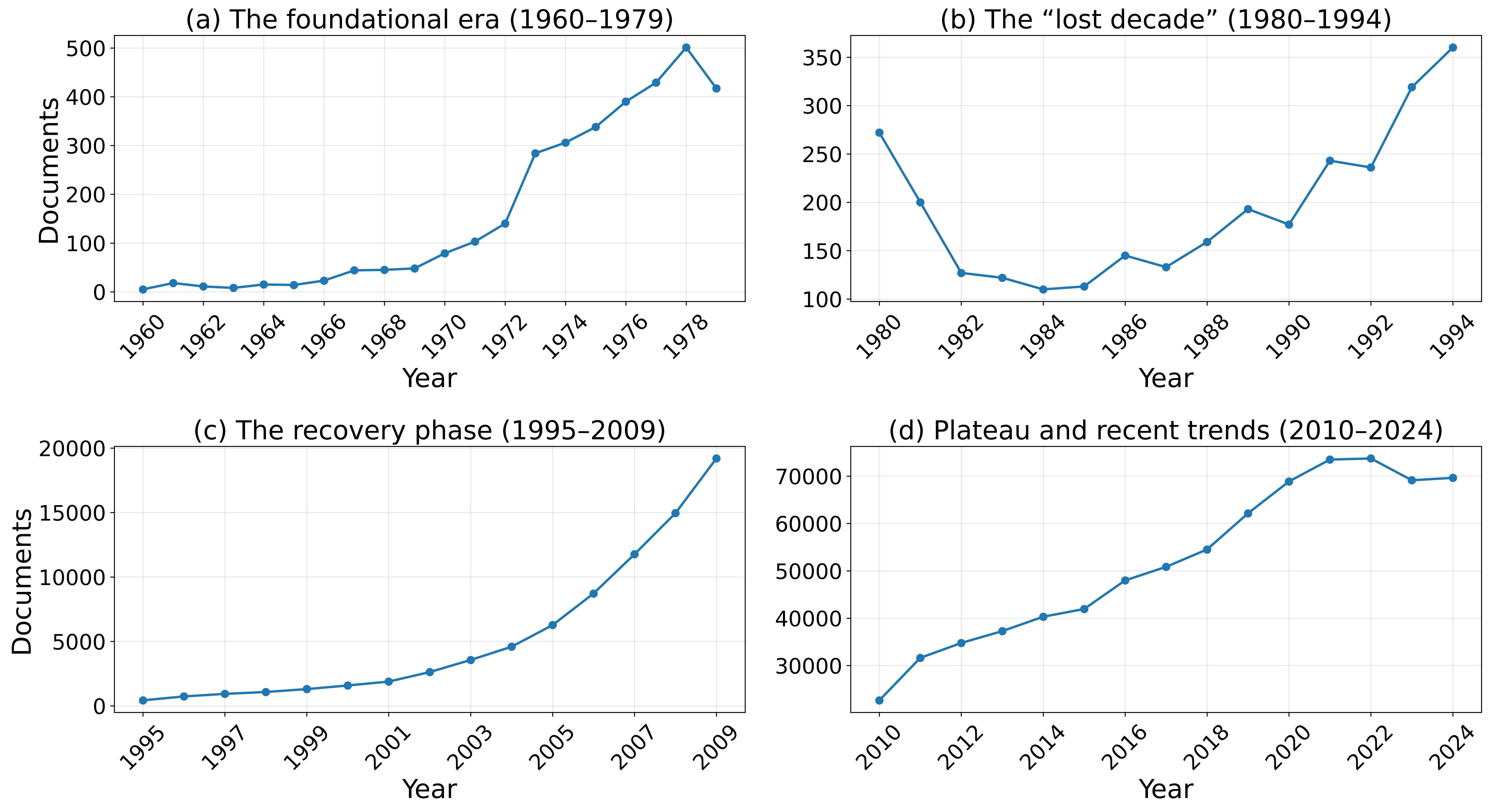} 
  \caption{Iran's publication output across four distinct periods, illustrating the pre-revolutionary growth, post-revolutionary collapse, subsequent recovery, and the recent plateau.}
  \label{fig:iran-periods}
\end{figure}

\begin{figure}[htbp]
  \centering
  \includegraphics[width=\textwidth]{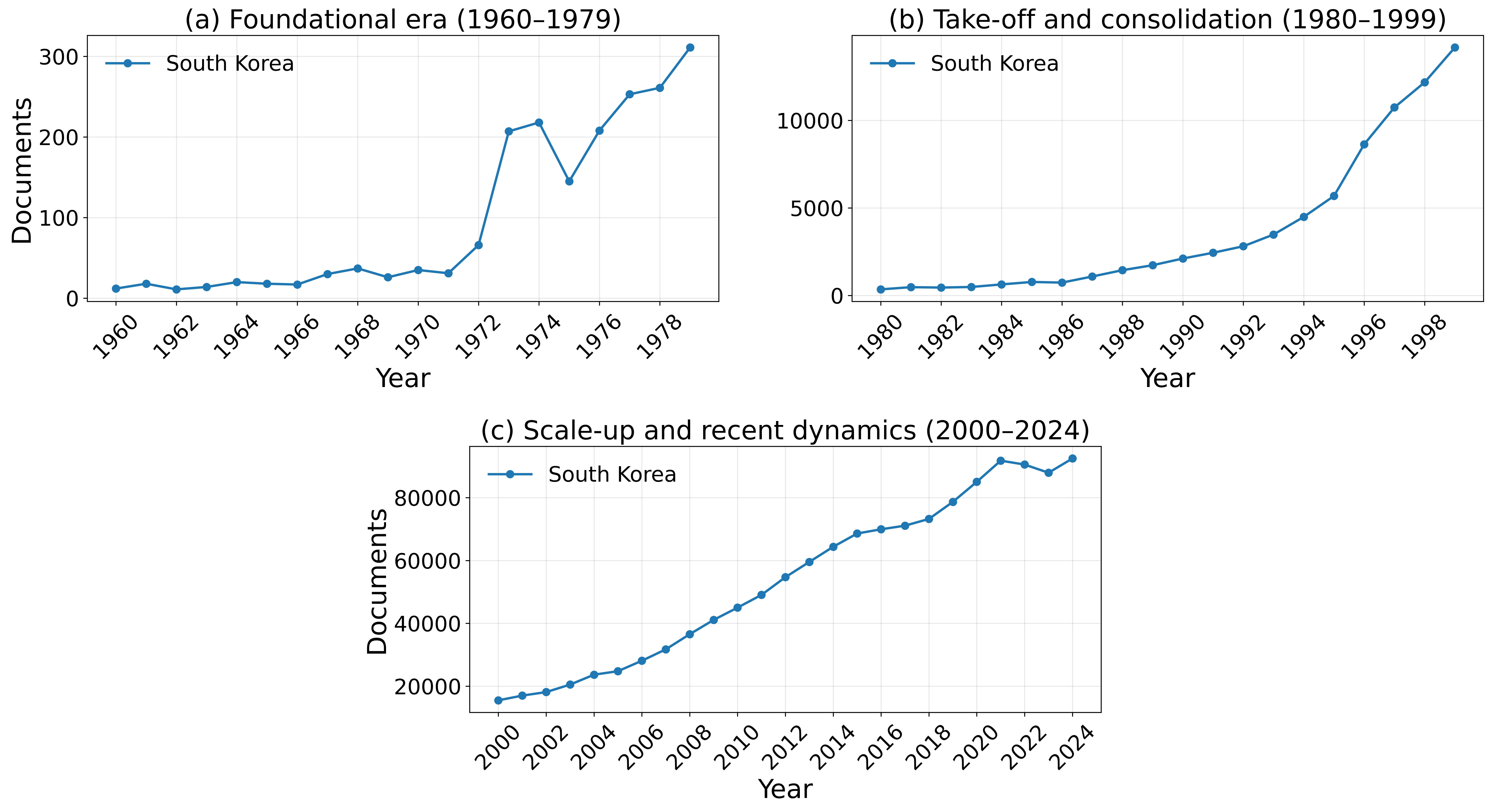} 
  \caption{Annual Scopus-indexed articles and reviews for South Korea (1960--2024) across three periods.}
  \label{fig:south-korea}
\end{figure}

\begin{figure}[htbp]
  \centering
  \includegraphics[width=\textwidth]{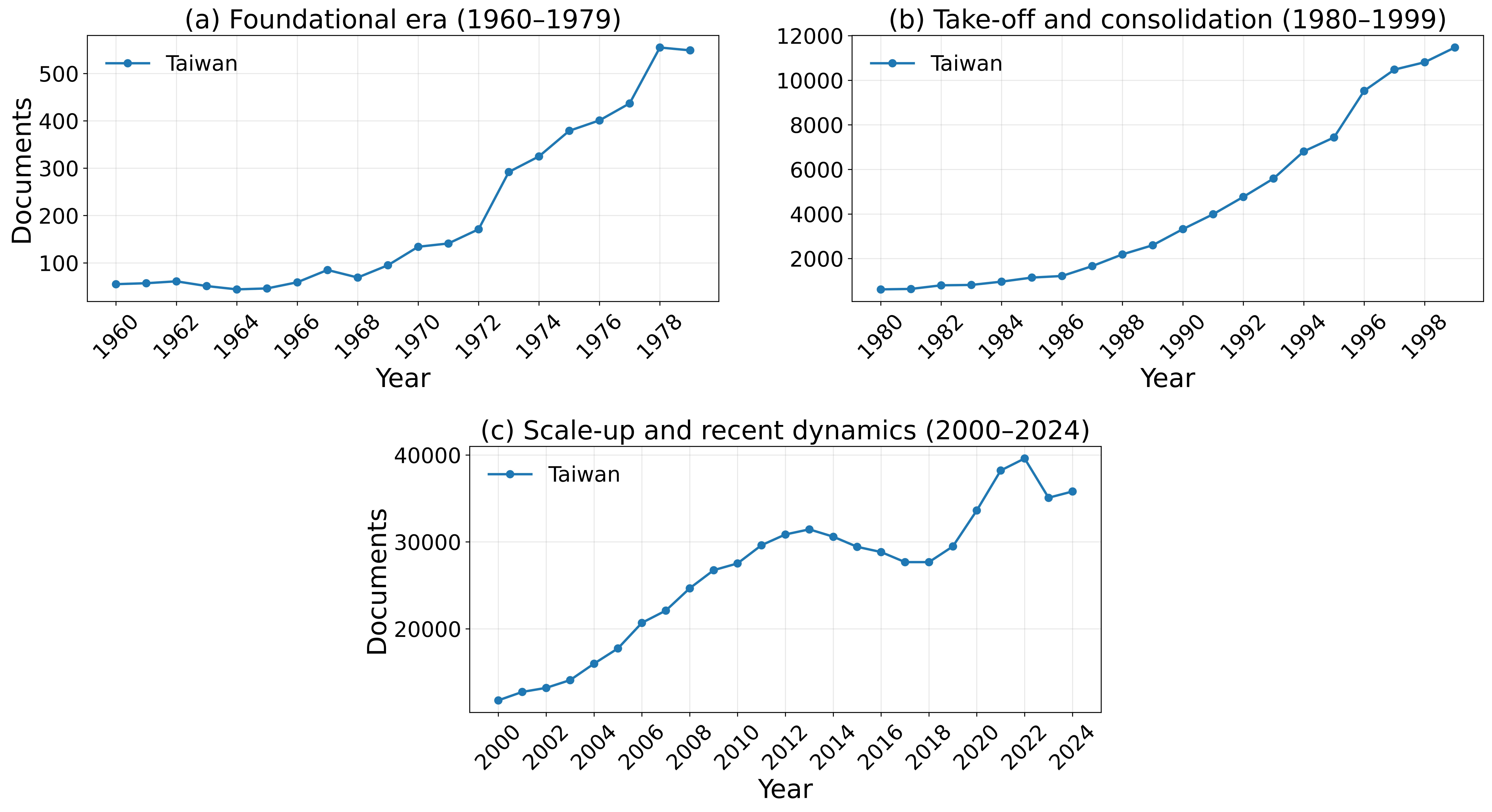} 
  \caption{Annual Scopus-indexed articles and reviews for Taiwan (1960--1979, 1980--1999, and 2000--2024).}
  \label{fig:taiwan}
\end{figure}

\begin{figure}[htbp]
  \centering
  \includegraphics[width=\textwidth]{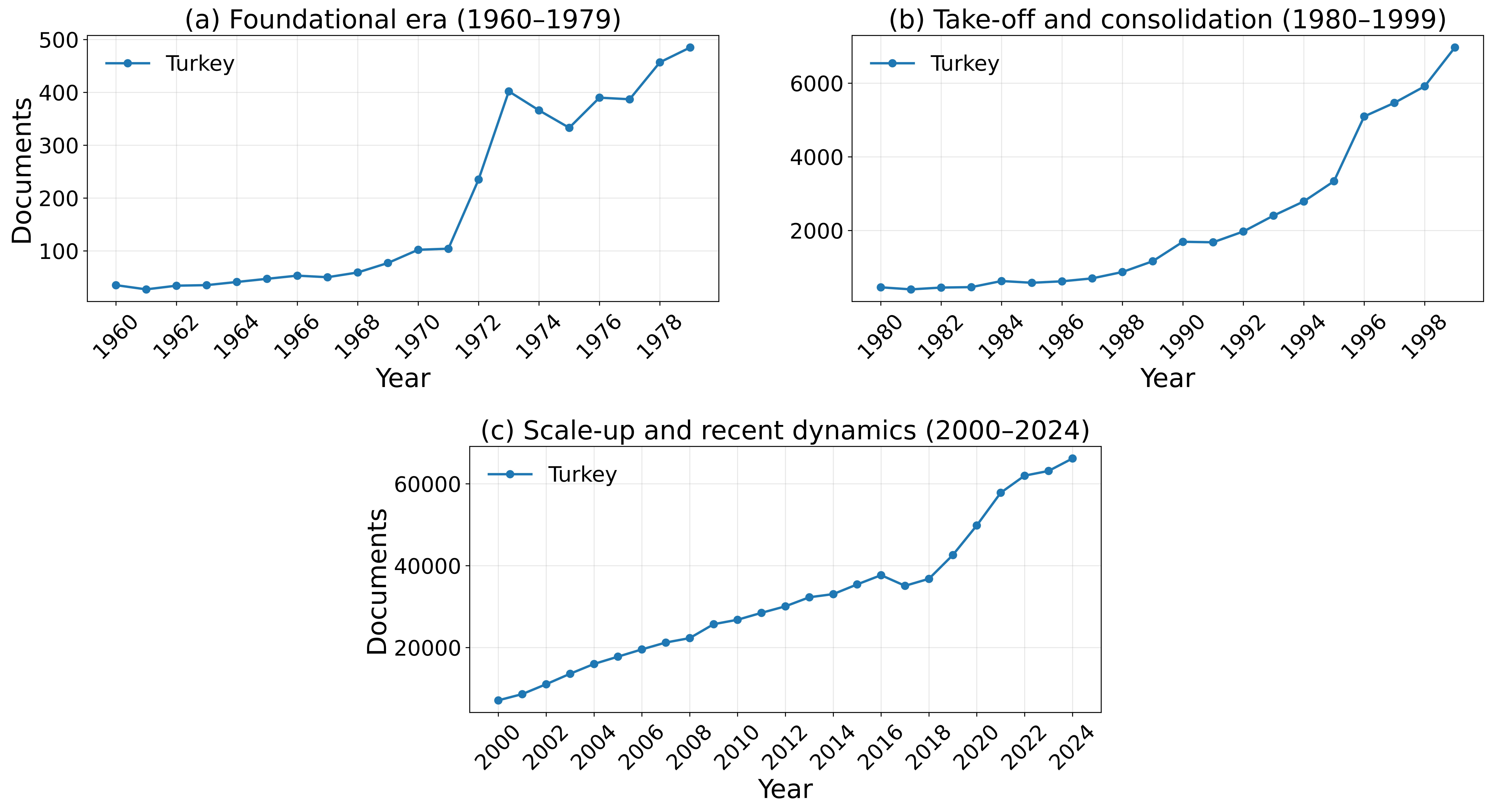} 
  \caption{Annual Scopus-indexed articles and reviews for Turkey (1960--1979, 1980--1999, and 2000--2024).}
  \label{fig:turkey}
\end{figure}

Our analysis reveals a story of divergence that unfolds not only in the volume of scientific output but also in its quality and international impact. We integrate publication counts with normalized citation and impact indicators to provide a multidimensional view of Iran’s scientific path in comparison with its Asian peers. 

\subsection{1960–1979: The Era of Iran’s Scientific Lead}

This period establishes the significant baseline: Iran was not a follower but an early leader among its Asian peers.  
Iran’s output grew rapidly in the 1970s, reaching 518 papers in 1978. In the same year:

- South Korea produced 260 papers,  
- China was at a comparable level, roughly in the 200–300 range,  
- Taiwan produced about 550 papers,  
- Turkey published approximately 470 papers by 1979.

Thus, Iran was clearly ahead of South Korea and roughly on par with Taiwan and Turkey, positioning it as a rising scientific player with strong upward momentum before 1979.

\subsection{1980–1999: The Great Divergence}

This 20-year period is where the trajectories of Iran and the Asian Tigers permanently diverged.

Iran’s output collapsed to a low of 122 papers in 1984 and did not surpass its 1978 peak until 1995. This period represents Iran’s ``lost decade,'' reflecting the combined effects of the Cultural Revolution’s university shutdowns and the reallocation of resources following the Iran–Iraq War.

In sharp contrast, South Korea, Taiwan, and China entered their take-off phase:

- South Korea grew from a few hundred papers in the early 1980s to over 17,000 papers in 1999,
- Taiwan expanded to more than 13,000 papers,
- China began a strong upward trajectory in the 1990s, surpassing 20,000 publications by the late decade,
- Turkey also accelerated, rising from a few hundred papers in the early 1980s to nearly 7,000 papers in 1999.

By the end of the century, the gap between Iran and its peers had evolved from a moderate separation into a wide, structural gap.  
This divergence was not only quantitative: the Asian Tigers were rapidly integrating into global scientific networks—boosting visibility, citation impact, and international collaboration—while Iran was facing geopolitical isolation and limited access to these networks.

\subsection{2000–2024: Recovery in Scale, Lag in Impact}

In the 21st century, Iran mounted a remarkable recovery in absolute output. Iran grew from about 2,500 papers in 2000 to a peak of roughly 80,000 papers in 2022. This rapid expansion reflects major national investments in mass higher education and research infrastructure.

However, Iran’s surge in quantity does not fully translate into gains in scientific impact.

Meanwhile, its peers followed different trajectories:

- China experienced an unprecedented rise, reaching more than 1,000,000 annual publications by 2024.
- South Korea surpassed 100,000 papers annually in recent years.
- Taiwan stabilized in the range of 50,000–60,000 papers.
- Turkey grew rapidly from about 7,000 papers in 2000 to more than 60,000 papers by 2024.

Despite Iran’s strong quantitative return, citation-based indicators show a persistent lag relative to the Asian Tigers. While South Korea, Taiwan, and Turkey have improved their field-normalized citation impact over time—and China has achieved near-parity with global averages in many fields—Iran remains significantly below these benchmarks.  
This pattern suggests a decoupling between scale and impact: Iran has achieved mass production, but its influence within the global scientific system expands at a much slower rate.

\subsection{Pre-Scopus Historical Reconstruction (1960–1995)}

To contextualize the Scopus-based results, Figures 5, 6 and 7 extend the analysis back to 1960 using Crossref’s metadata and citation-to-date records.
Because Scopus coverage before the mid-1990s is incomplete, Crossref provides a valuable
approximation of earlier publication and citation trends.

\begin{figure}[htbp]
  \centering
  \includegraphics[width=\textwidth]{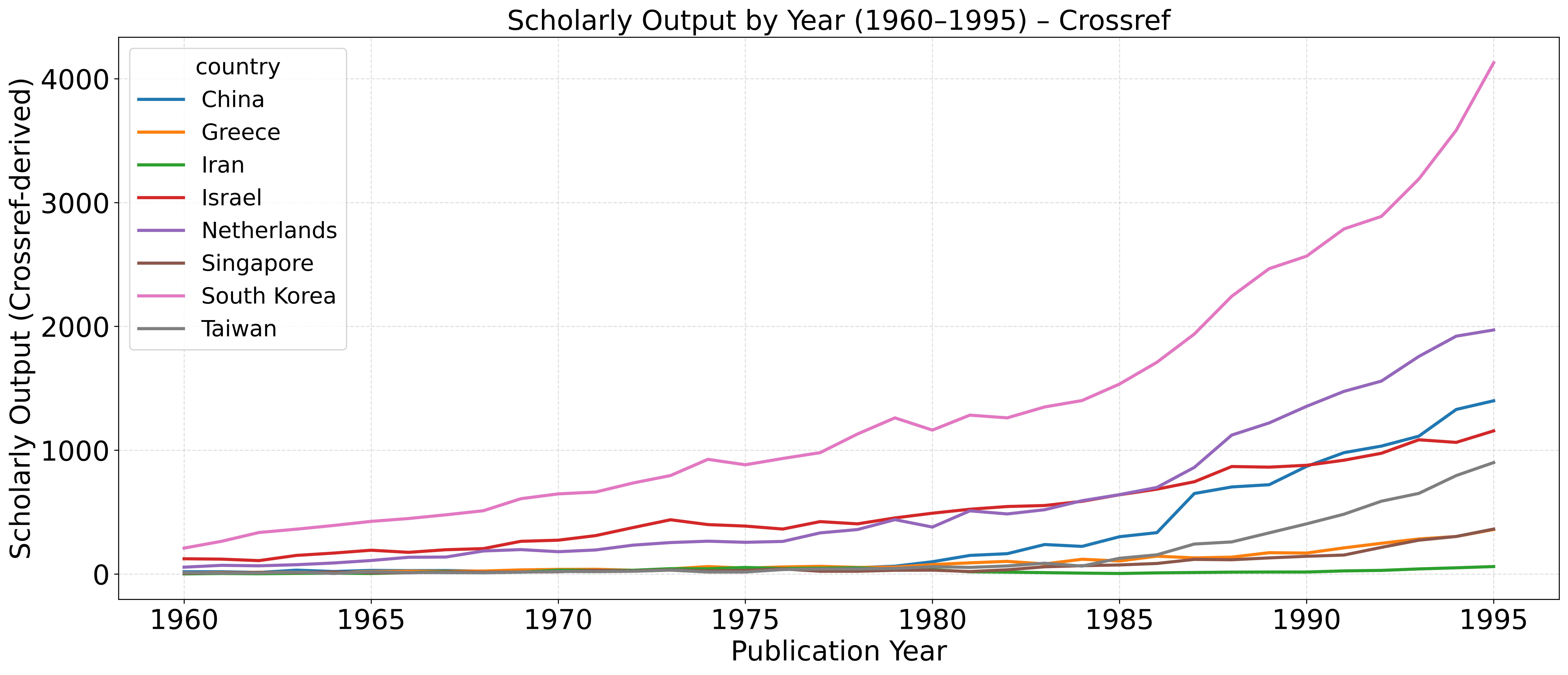} 
  \caption{Annual scholarly output (1960--1995) based on Crossref data.}
  \label{fig:crossref-output}
\end{figure}

\begin{figure}[htbp]
  \centering
  \includegraphics[width=\textwidth]{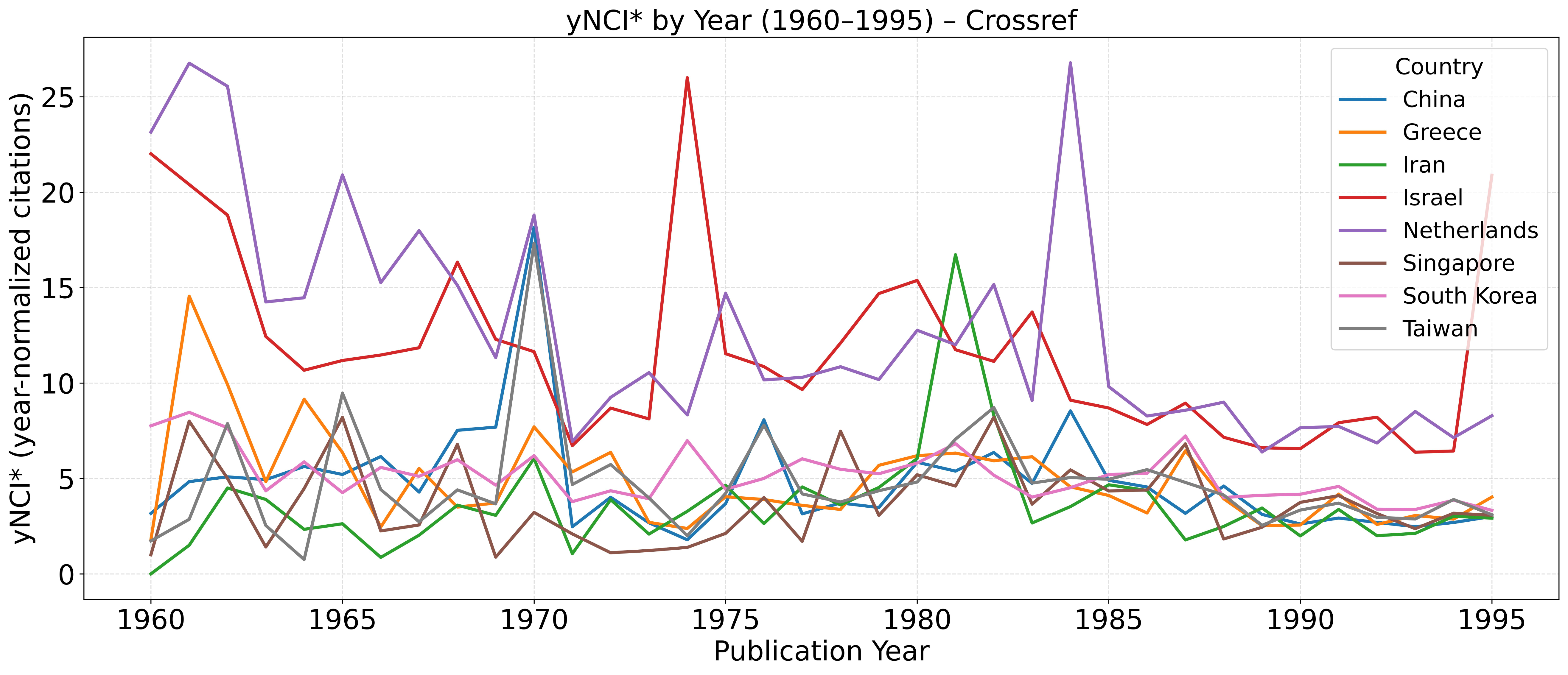} 
  \caption{Approximate year-normalized citation index (yNCI*) for 1960--1995, reconstructed from Crossref citation records.}
  \label{fig:crossref-ynci}
\end{figure}

\begin{figure}[htbp]
  \centering
  \includegraphics[width=\textwidth]{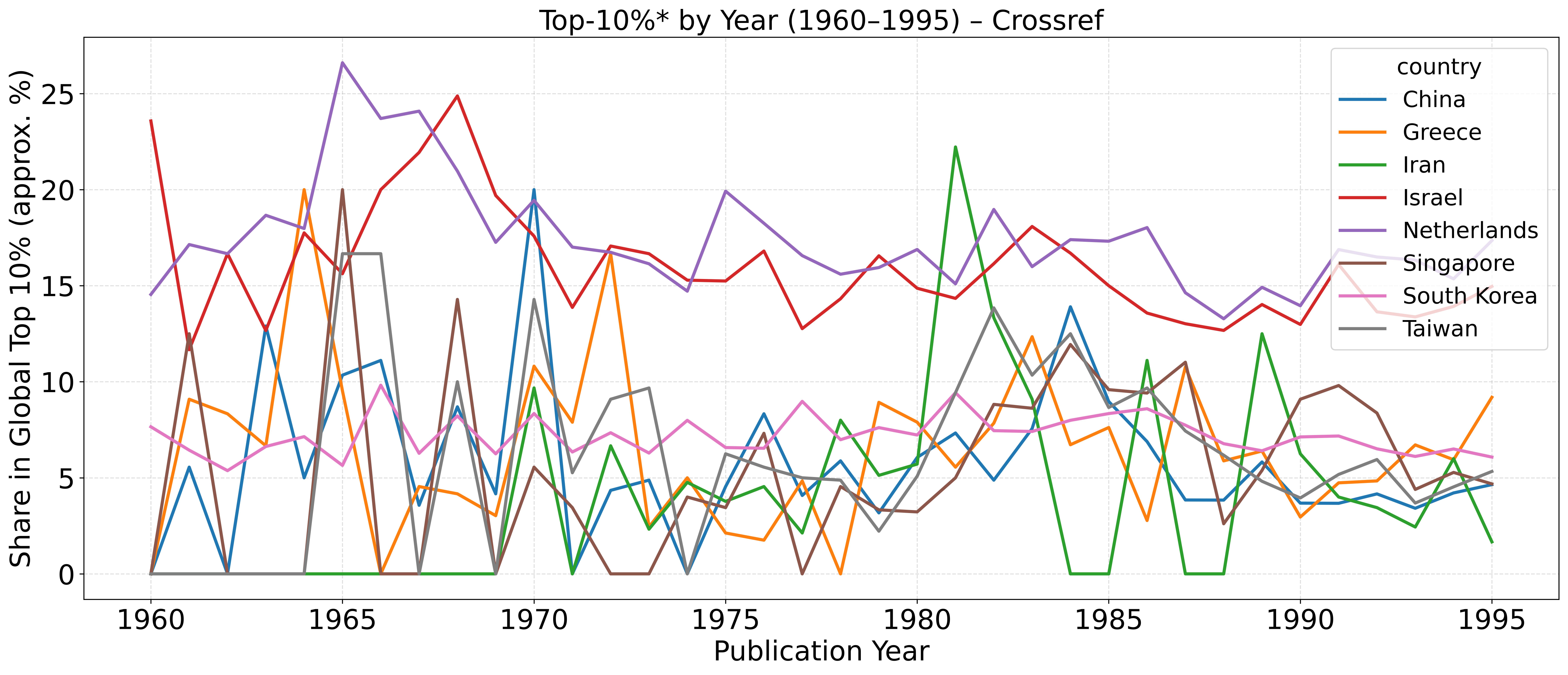} 
  \caption{Share of publications in the approximate global top 10\%* citation percentile (Top-10\%*) between 1960--1995.}
  \label{fig:crossref-top10}
\end{figure}

These reconstructed metrics—denoted by an asterisk (\textit{yNCI*}, \textit{Top-10\%*})\-/use
total citation counts to date rather than fixed five-year windows, offering a reasonable
historical baseline for comparing national trajectories prior to the 1979 Revolution.
The patterns corroborate the main narrative: Iran’s pre-1979 scientific system exhibited
healthy compounding growth and moderate global visibility, both of which collapsed
abruptly during the 1980s before partial recovery in the 1990s.

\subsection{The Paradox of Quantity vs. Quality}

Figures 8, 9 and 10 illustrate the central paradox of Iran's post-2000 recovery: a successful restoration of research \textit{scale} (quantity) alongside a persistent \textit{lag} in research \textit{influence} (quality).

\begin{figure}[htbp]
  \centering
  \includegraphics[width=\textwidth]{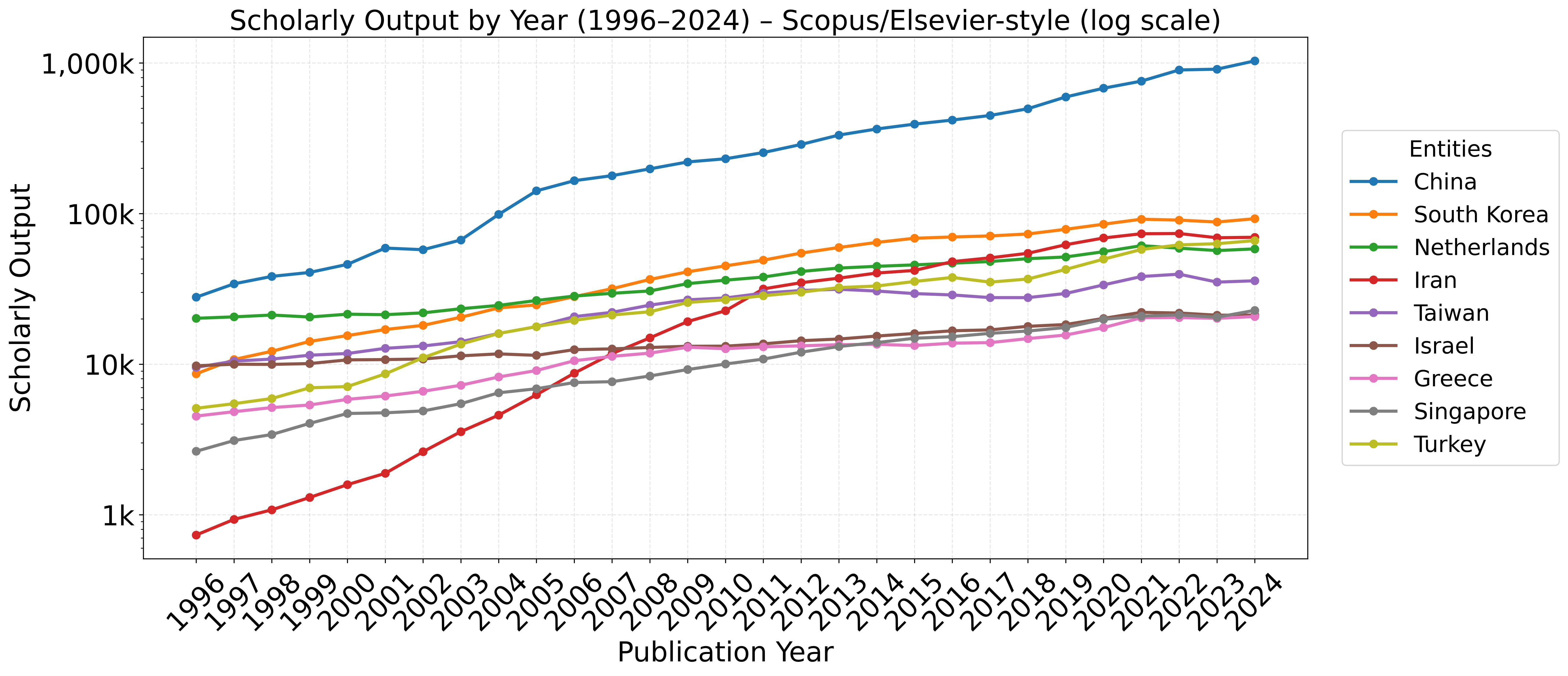} 
  \caption{Annual scholarly output of Iran and selected countries (1996--2024).}
  \label{fig:scival-output}
\end{figure}

\begin{figure}[htbp]
  \centering
  \includegraphics[width=\textwidth]{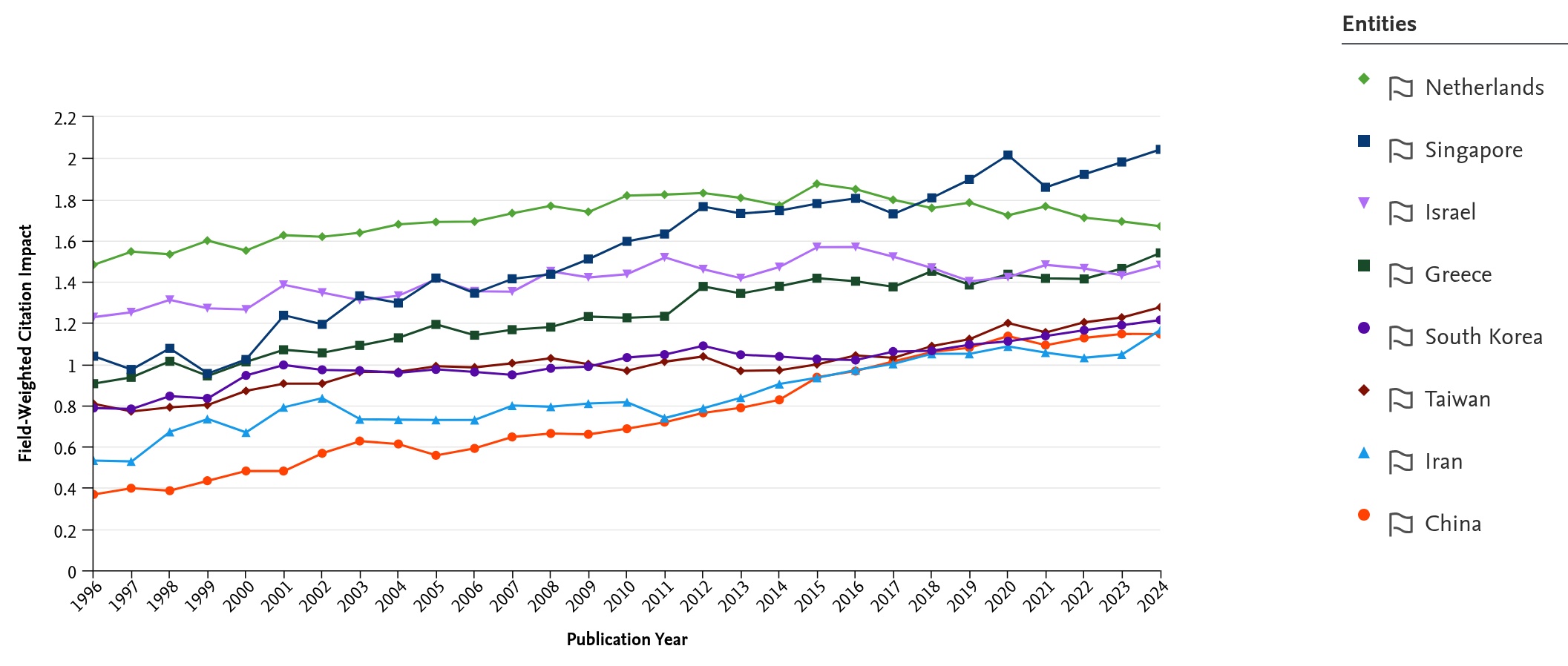} 
  \caption{Field-Weighted Citation Impact (FWCI) by publication year (1996--2024).}
  \label{fig:fwci}
\end{figure}

\begin{figure}[htbp]
  \centering
  \includegraphics[width=\textwidth]{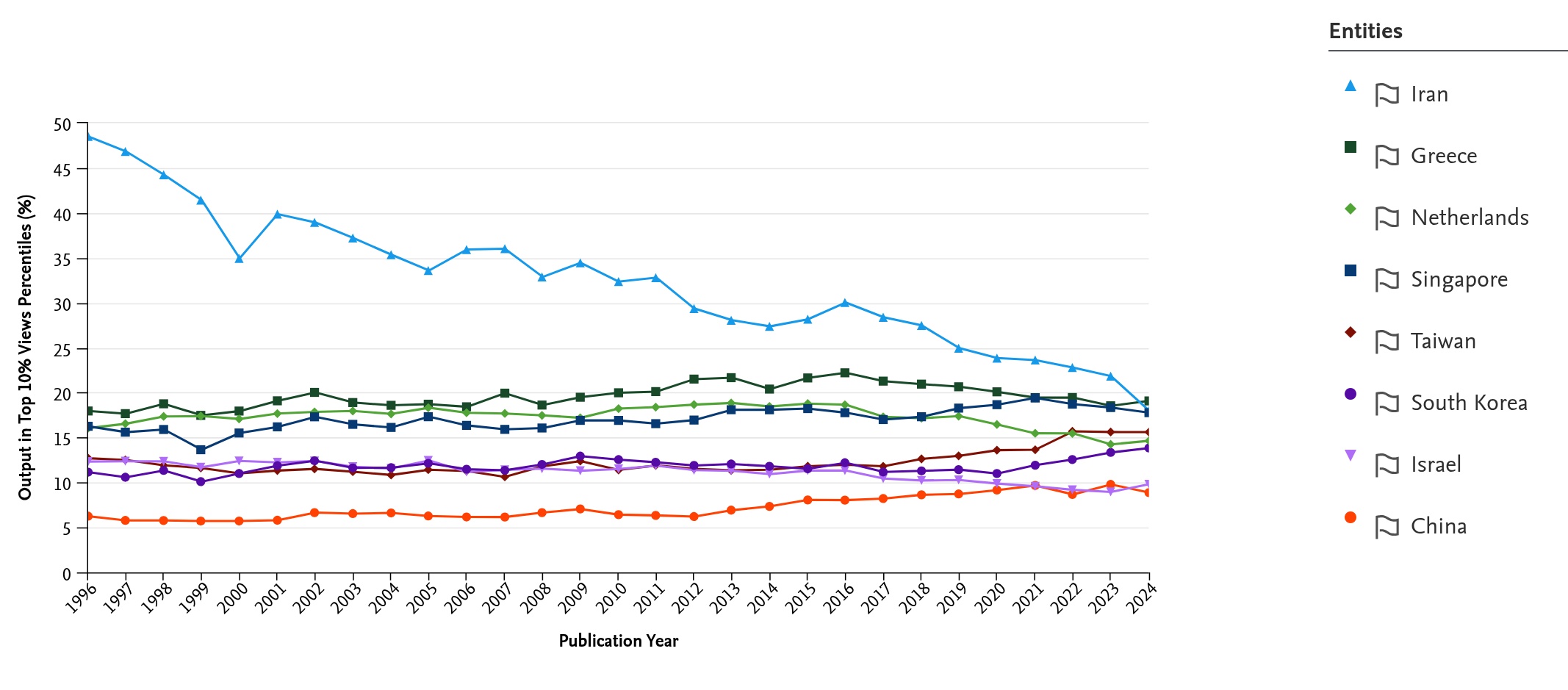} 
  \caption{Percentage of publications within the Top 10\% citation percentiles (1996--2024).}
  \label{fig:top10-scival}
\end{figure}

The combined trends in Figure 8 reveal that Iran’s research output expanded sharply after the early 2000s, demonstrating a remarkable recovery in absolute numbers. However, this growth in scale obscures a significant quality deficit when analyzed with normalized metrics.

As shown in Figure 9, Iran's Field-Weighted Citation Impact (FWCI)—a measure of citation impact normalized for subject field, year, and document type—showed gradual improvement, rising from well below the world average to approach it ($\text{FWCI} \approx 1.0$) by 2020. While a significant achievement, this stands in stark contrast to its high-performing peers. Elite scientific nations like the Netherlands and Singapore consistently maintained an FWCI above 1.5, meaning their research was, on average, 50 percent more cited than the global average. Even South Korea and Taiwan sustained an FWCI comfortably above 1.2.

A similar trend is visible in the production of elite science. Figure 10 shows the percentage of publications in the top 10 percent most-cited papers globally. Iran's share grew impressively from under 5 percent in the late 1990s to around 12 percent by 2024. Yet, this figure still lags behind top-tier systems like the Netherlands and Singapore, which consistently place over 25 percent of their output in this elite category.

This persistent gap between quantity and quality, while common in rapidly expanding systems, is particularly pronounced in the Iranian case. We argue this lag is not a temporary phase but the result of enduring structural factors, operating at both the international and domestic levels.

Internationally, the mechanisms are clear. Decades of relative "isolation, exacerbated by geopolitical constraints and sanctions", have created a structural limitations on Iranian science. This extends beyond just constraining travel or funding; it has limited opportunities for "deep collaboration with leading global research groups". As international collaboration is a primary driver of citation impact (FWCI), this "reduced integration into elite networks inherently limits knowledge transfer and citation visibility". This isolation has meant limited access to the latest equipment, shared data, and the informal knowledge transfer that drives scientific frontiers, thereby constraining the potential impact of the research being produced.

Domestically, the drivers of the quality gap are equally significant and are, paradoxically, linked to the quantitative recovery itself. The state-led push to rebuild the scientific enterprise from the late 1990s onward was heavily influenced by a desire to reclaim national prestige through global university rankings. This created a "science policy and university promotion criteria" system that "often prioritized quantitative metrics over impact". This incentive structure created a rational response: "volume maximization strategies". This includes "publishing in lower-tier venues or engaging in incremental research" to meet promotion quotas, often "at the expense of the long-term, resource-intensive projects that typically yield high-impact breakthroughs".

Consequently, the very policies that fueled the impressive recovery in \textit{scale} (Figure 8) may have inadvertently entrenched the deficit in \textit{influence} (Figures 9 and 10). To synthesize these multi-dimensional findings, we present a comparative profile in Table 1.

\begin{table}[h!]
\centering
\caption{Comparative Scientometric Profile of Iran and Peer Group (1996-2024)}
\label{tab:quality_metrics}
\begin{tabular}{lcccc}
\hline
\textbf{Country} & \textbf{Total Output} & \textbf{Avg. FWCI} & \textbf{Avg. Top 10\%} \\
& \textbf{(1996-2024)} & \textbf{(2015-2024)} & \textbf{(\%)} \\
\hline
Iran & $\approx 750k$ & $\approx 1.0$ & $\approx 12\%$ \\
South Korea & $\approx 1.2M$ & $\approx 1.2$ & $\approx 15\%$ \\
Taiwan & $\approx 600k$ & $\approx 1.2$ & $\approx 15\%$ \\
Singapore & $\approx 350k$ & $>1.5$ & $>25\%$ \\
Netherlands & $\approx 1.1M$ & $>1.5$ & $>25\%$ \\
Israel & $\approx 450k$ & $\approx 1.4$ & $\approx 20\%$ \\
\hline
\end{tabular}
\end{table}

\subsection{Cross-country synthesis on a log scale}
The comparative analysis in Figure 11, which juxtaposes trajectories on a logarithmic scale, provides the clearest visualization of the divergence. It reveals how small differences in growth \textit{rates} (slopes), when sustained over decades, compound into massive, permanent differences in \textit{level}. The vertical dashed line marks the 1979 breakpoint.

\begin{figure}[htbp]
  \centering
  \includegraphics[width=\textwidth,height=0.8\textheight,keepaspectratio]{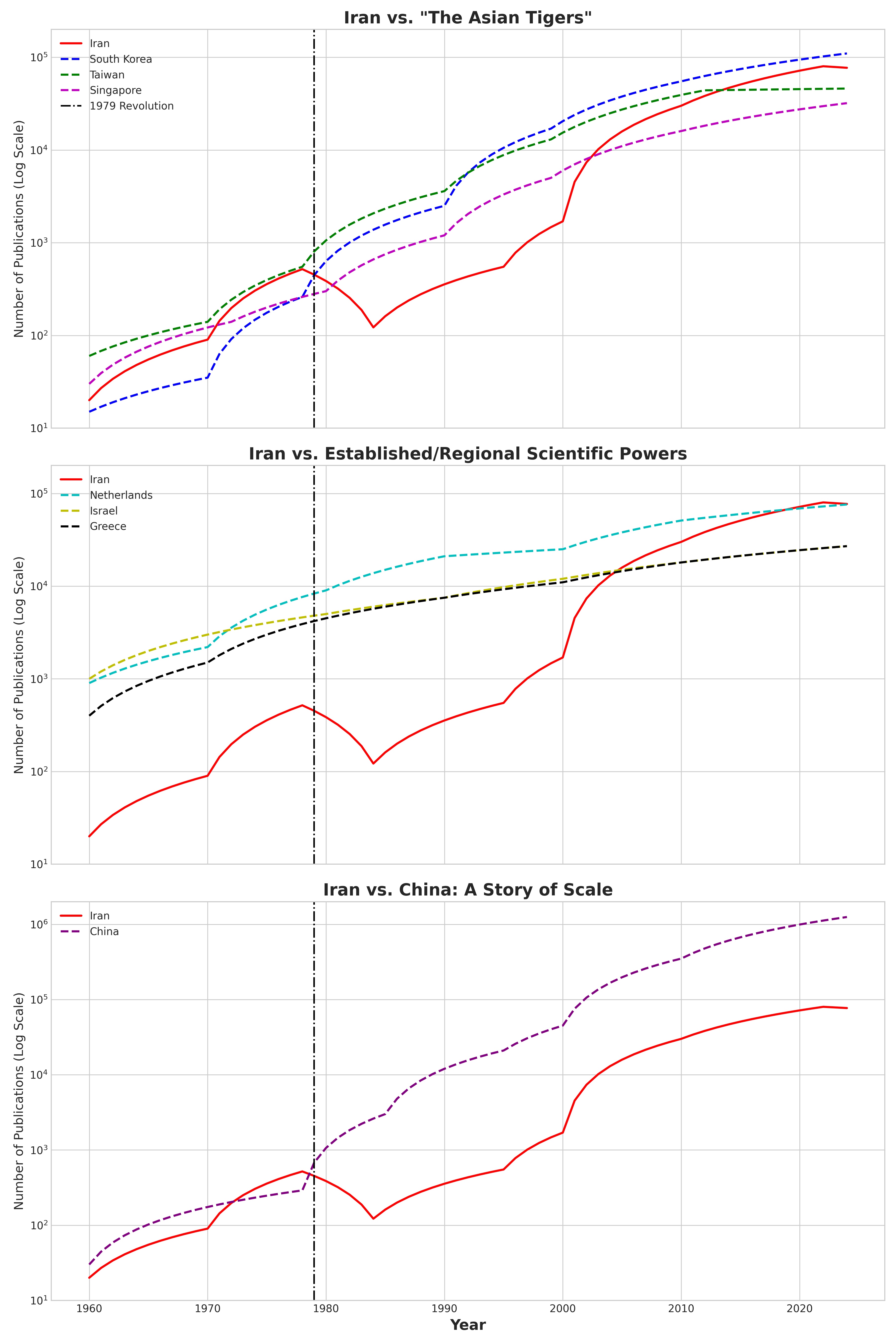}
  \caption{Comparative growth trajectories for Iran and its peer groups.}
  \label{fig:traj-peers}
\end{figure}

Across all panels, the data demonstrates that (i) Iran’s pre-1979 slope was competitive, (ii) the post-1979 regime shift dramatically flattened this slope, raising its implied doubling time for roughly a decade and a half, and (iii) peers that maintained steep, low-doubling-time regimes (the Tigers; the Netherlands/Israel; later China) compounded into permanently higher levels. The resulting divergence is thus a consequence of \emph{rate differentials sustained over long horizons}, not merely one-off level shocks.

This divergence in compounding growth is evident across all peer groups. In the comparison with the "Asian Tigers" (Panel A), Iran’s pre-1979 momentum is clear; its slope is competitive—at times steeper than South Korea’s—and its level is comparable to Taiwan’s. Immediately after 1979, Iran’s curve flattens and dips, signalling a collapse in compounding. By contrast, South Korea, Taiwan, and Singapore enter multi-decadal high-slope regimes during the 1980s–1990s. This fundamental divergence in \emph{rates} explains why Iran’s strong 2000s recovery does not translate into convergence.

Similarly, when compared to established systems (Panel B), the Netherlands and Israel trace mature, high-level trajectories with moderately rising slopes, consistent with large, stable research systems. Relative to this group, Iran’s post-1979 dip is distinctive: the level falls below Greece in the 1980s and only re-approaches the regional band after 2000. The key contrast is regime stability: the mature systems preserve low implied doubling times throughout the 1980s–1990s, whereas Iran shifts into a high-doubling-time regime.

Finally, the comparison with China (Panel C) illustrates a bifurcation of paths. While Iran experiences a trough and slow rebuild, China undergoes a two-stage regime shift: a steepening in the 1990s and an even steeper rise post-2000. The persistent slope gap through the 1990s–2010s makes absolute catch-up arithmetically implausible for Iran, even with its robust 2000s growth.

\subsection{Per-capita capacity}
Normalizing by population (Figure 12) clarifies that the Asian Tigers exhibit early inflection and multi-decadal compounding on a per-capita basis, reflecting institutional deepening (graduate training, research careers, internationalization). Iran’s post-1979 trough is stark on this metric: per-capita output falls sharply, then climbs in the 2000s, but the \emph{relative} gap with Tigers and mature systems (Netherlands, Israel) remains large. The normalization shows Iran’s 2000s recovery was driven primarily by scale rebuilding rather than a per-capita surge sufficient for convergence.

\begin{figure}[htbp]
  \centering
  \includegraphics[width=\textwidth,height=0.8\textheight,keepaspectratio]{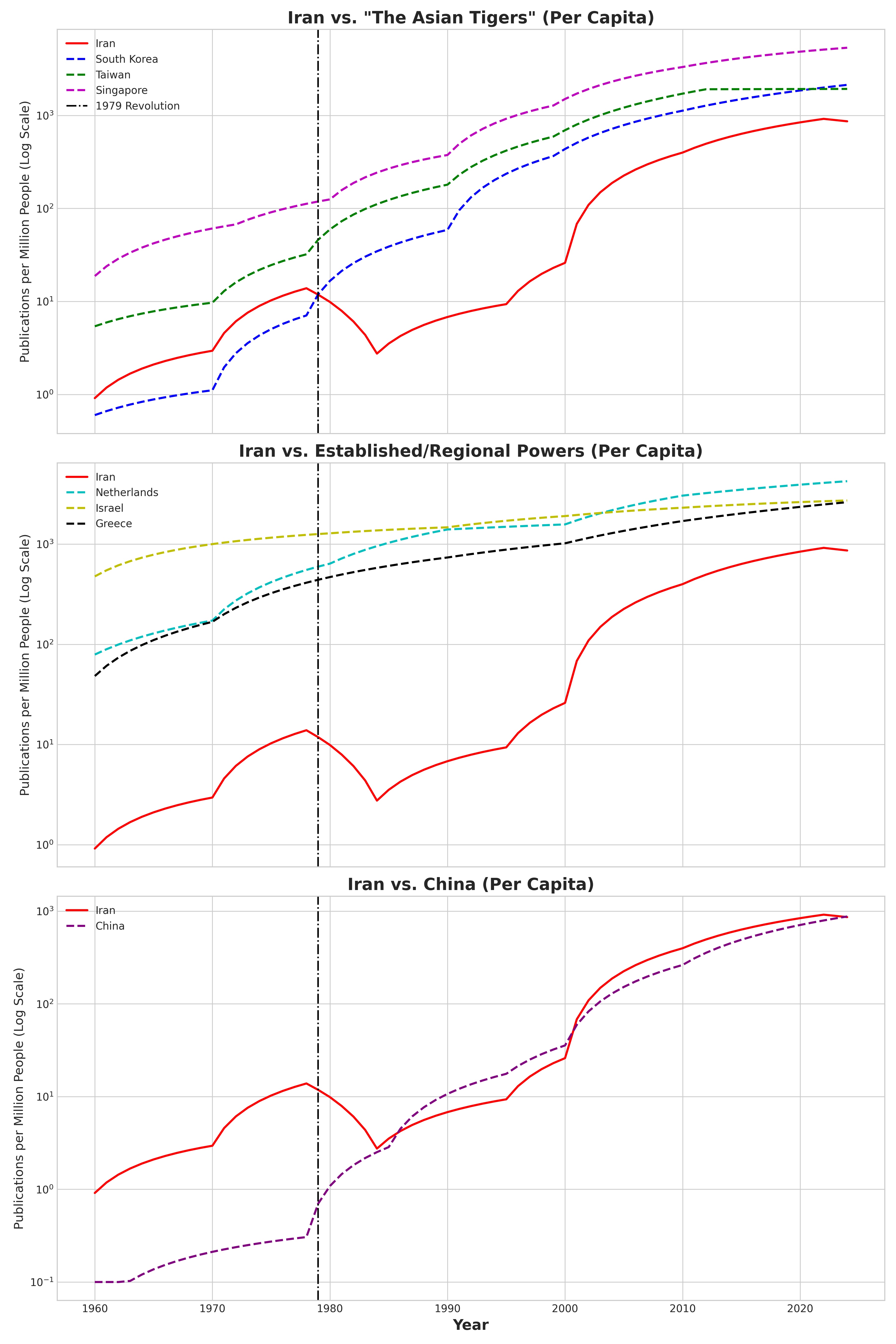}
  \caption{Publications per million people (log scale). Normalization by population reveals capacity per person. Iran's post-1979 dip and delayed catch-up contrast with the continuous climb of the Asian Tigers and mature systems.}
  \label{fig:per-capita}
\end{figure}

\subsection{Scientific growth momentum: implied years to double output}
The Figure tracks \emph{implied years to double} publication output---a momentum metric computed from rolling growth rates (lower values indicate stronger compounding). The vertical dashed line marks 1979.

\begin{figure}[htbp]
  \centering
  \includegraphics[width=\textwidth]{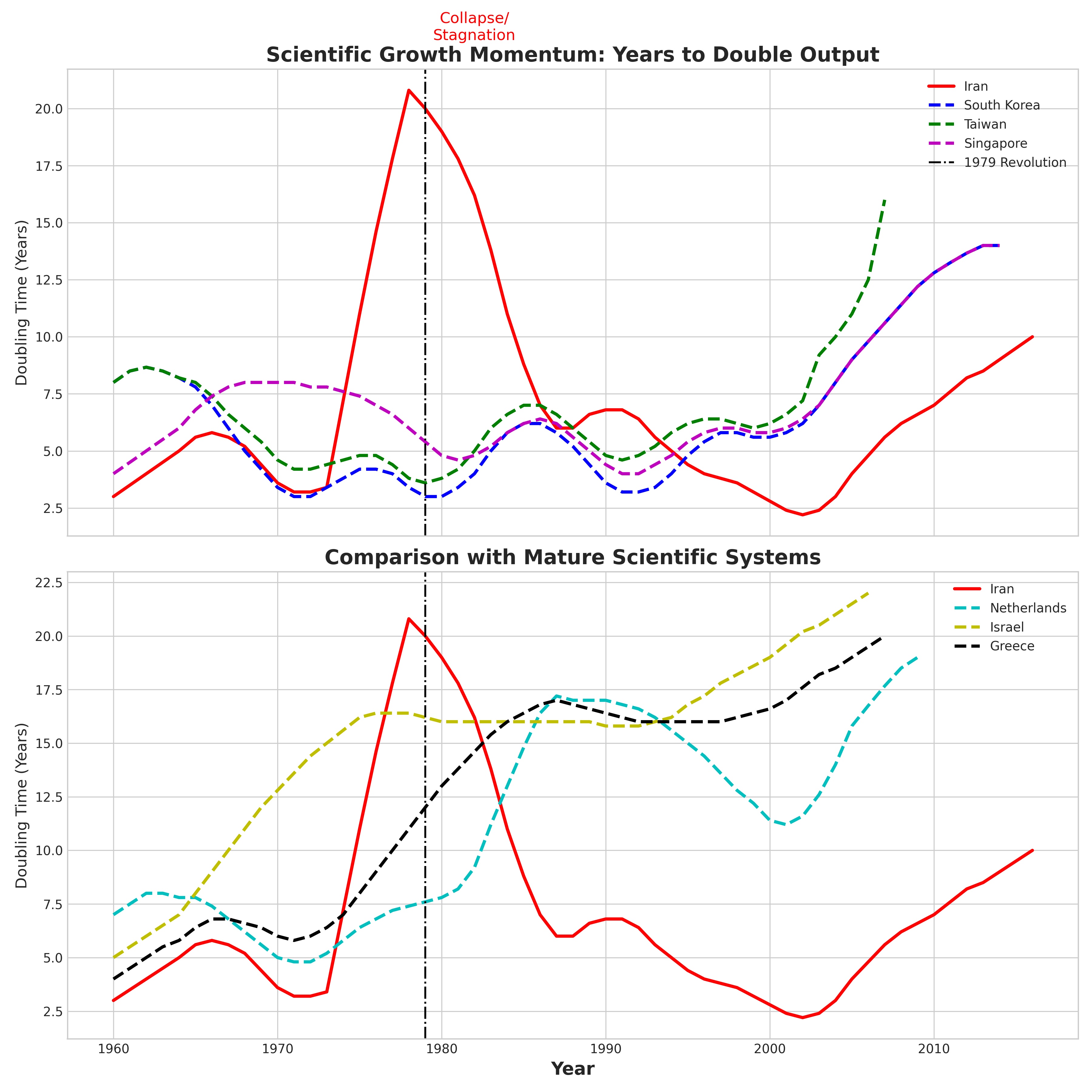}
  \caption{Growth momentum and implied doubling times.}
  \label{fig:doubling-time}
\end{figure}

\paragraph{Panel A (top): Iran vs. the Asian Tigers.}
Iran enters the 1970s with comparatively \emph{low} doubling times (about 3--5 years), signalling an active, fast-growing system. Around 1979--1981 the red curve spikes abruptly to well above 20 years, a signature of stagnation; this regime shift coincides with the revolution and its immediate sequelae. Momentum then gradually normalizes in the late 1980s (back toward \(\sim\)6--7 years), briefly improves around 2000 (approaching \(\sim\)3 years), and subsequently deteriorates again during the 2010s (rising toward \(\sim\)8--10 years).

In sharp contrast, South Korea and Taiwan maintain \emph{persistently low} doubling times through the 1980s--1990s (roughly \(\sim\)3--6 years), exactly the regime required for multi-decadal exponential scale-up. Singapore shows a similar pattern: momentum strengthens through the 1980s--1990s, then lengthens in the 2010s as a maturing system naturally slows. Panel~A shows that the Tigers \emph{sustain} low doubling times for long horizons just as Iran transitions into a high-doubling-time regime circa 1979; even Iran’s later recovery does not keep momentum low for long enough to deliver convergence.

\paragraph{Panel B (bottom): Iran vs. mature/regional scientific systems.}
The Netherlands and Israel display the characteristic profile of \emph{mature} research ecosystems: doubling times hover in the low-teens and gradually lengthen over time, reflecting steady compounding from a high base rather than explosive growth from a low base. Greece climbs from mid-single-digit doubling times in the 1960s toward the teens by the 1990s and 2000s, consistent with long-horizon capacity building in a smaller system. Against this backdrop, Iran’s trajectory is distinctive: a pre-1979 low-doubling-time regime (fast momentum), a dramatic spike to \(>\)20 years at the turning point, a late-1980s normalization, a short early-2000s before slowing again.

\subsection{Counterfactual Analysis: Quantifying the Knowledge Deficit}\label{Counterfactual Analysis}


To quantify the magnitude of the divergence, we developed a suite of counterfactual models, each designed to project Iran's scientific output under a scenario of uninterrupted growth. These models range from simple mathematical extrapolations to more sophisticated, empirically grounded simulations. The results, visualized in Figure 14, collectively illustrate the significant opportunity cost of the post-revolutionary disruption.

\begin{figure}[htbp]
  \centering
  \includegraphics[width=\textwidth]{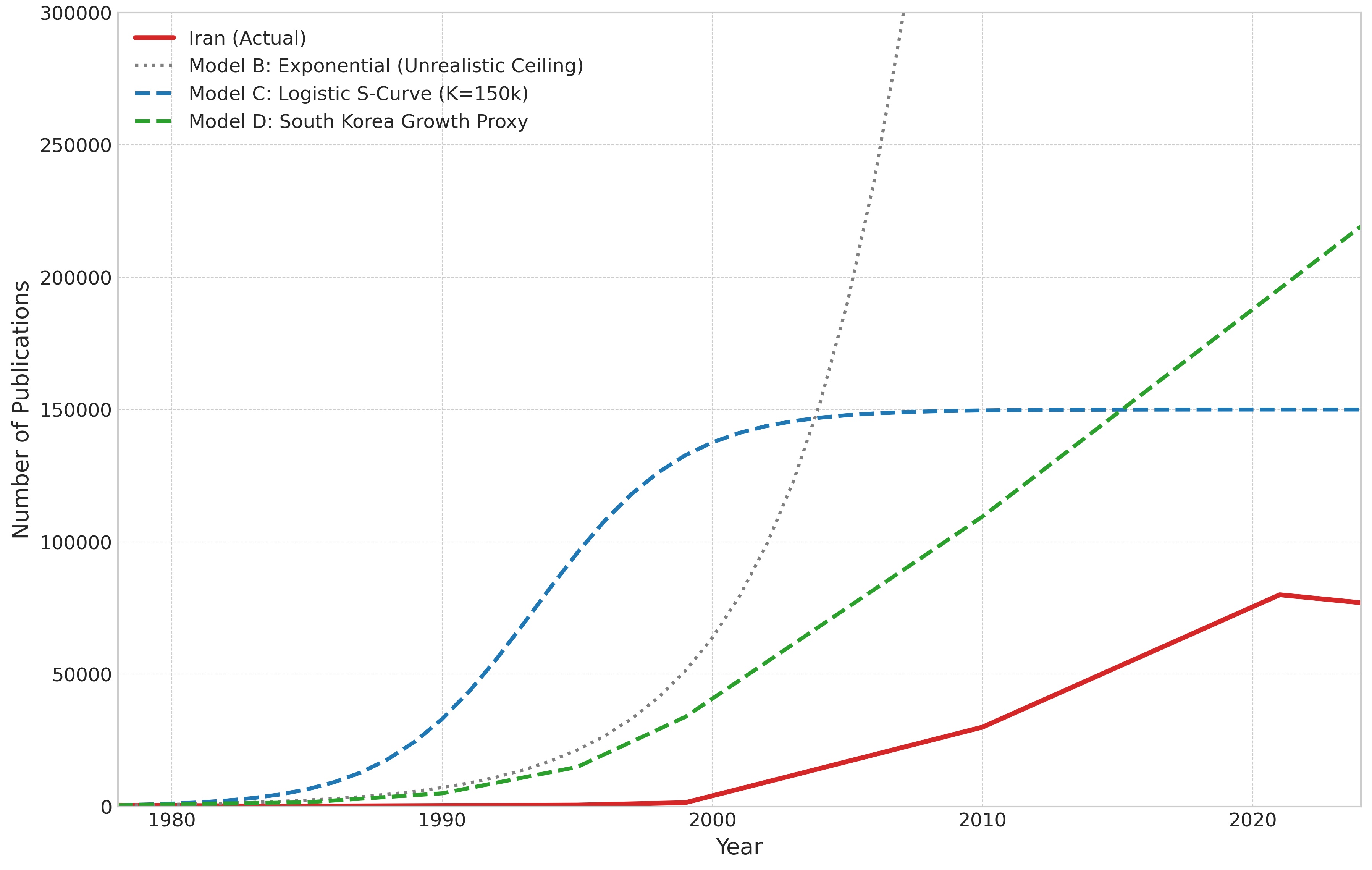}
  \caption{Counterfactual projections for Iran's scientific output. The solid red curve shows the actual trajectory. Model B (grey, dotted) is an exponential upper bound (unrealistic ceiling). Model C (blue, dashed) follows a logistic S-curve with carrying capacity $K=150{,}000$. Model D (green, dashed) applies South Korea's historical year-on-year growth rates to Iran's 1978 base.}
  \label{fig:counterfactual-models}
\end{figure}

\subsubsection{Model B (Perpetual Exponential Growth)}
This model serves as a theoretical benchmark to illustrate the power of compounding. It assumes that the institutional momentum Iran had built up in the decade prior to the revolution would continue indefinitely. We calculate the Compound Annual Growth Rate (CAGR) from 1970 to 1978, which stood at a robust $r \approx 13.1\%$. The projection is then calculated using the standard exponential growth formula:
$$\hat{N}_{IR,t}^{B} = N_{IR,1978} \times (1+r)^{(t-1978)}$$
As shown by the dotted gray line in Figure 14, this model projects an unrealistic output of over 1.2 million papers by 2024. While no system can sustain such a high growth rate forever, it highlights a significant mathematical limitation for Iran's lost potential, suggesting a trajectory that could have theoretically rivaled the scale of modern-day China.

\subsubsection{Model C (Logistic S-Curve Growth with Carrying Capacity)}
This model introduces a significant element of realism: the concept of maturation. Scientific systems, like all growth processes, eventually face diminishing returns as they exhaust available resources (e.g., funding, human capital, institutional capacity), causing their growth to slow and follow an S-shaped curve. The logistic model captures this by incorporating a "carrying capacity" ($K$), which represents the plausible upper limit of a nation's scientific system.

\textbf{Justification for Carrying Capacity (K):} We set $K=150,000$ publications per year. This figure is not arbitrary; it is chosen to represent the scale of a mature, top-tier, and stable scientific nation, comparable to the current output of countries like South Korea, Italy, or Canada. In the context of scientometrics, K is a function of sustained investment, population size, and effective science policy. Nations achieving this scale of output typically maintain a Gross Expenditure on R\&D (GERD) consistently above 2\% of GDP, coupled with a high density of researchers within the population. The value of 150,000 thus represents a plausible limitation that a nation with Iran's demographic potential and pre-1979 momentum could have realistically achieved, had it maintained stability and adopted similar long-term investment strategies.

The logistic formula is:
$$\hat{N}_{IR,t}^{C} = \frac{K}{1 + \left(\frac{K}{N_{1978}}-1\right)e^{-r(t-1978)}}$$
The dashed blue line in Figure 14 shows this path. Iran's growth is initially rapid, but then subsequently slows as it approaches the 150,000-publication mark. This model projects a far more plausible scenario where Iran would have become a major, stable scientific power.

\subsubsection{Model D (South Korea as a Growth-Proxy)}
This final model is compelling as it is grounded in a real-world success story. It answers the question: "What if Iran, starting from its superior position in 1978, had simply followed South Korea's subsequent path of development?"

\textbf{Justification for Proxy Selection:} The choice of South Korea as a proxy is theoretically motivated. In comparative research, suitable proxies are those that share critical baseline characteristics but differ on the key variable of interest (in this case, political stability post-1979). Prior to 1979, both Iran and South Korea were on a similar trajectory of state-led modernization and industrialization, with science and technology policy identified as a central pillar of national development. South Korea's subsequent success represents the archetypal model of a "developmental state" that maintained political stability and executed a long-term, strategic vision for building a knowledge-based economy. Therefore, its growth trajectory serves as a credible "alternate timeline" for a nation that started from a similar baseline but did not experience a disruptive political shock.

Instead of a fixed rate, this model applies the actual, historical year-on-year growth percentage of South Korea $(g_{SK,t})$ to Iran's trajectory from 1979 onwards:
$$\hat{N}_{IR,t}^{D} = \hat{N}_{IR,t-1}^{D} \times (1+g_{SK,t})$$
The result, as shown by the dashed green line in Figure 14, indicates that Iran would not only match but significantly surpass South Korea, reaching an annual output of approximately 245,000 papers by 2024. This is because Iran would have been applying Korea's high growth rates to its own, larger initial base of scientific output. This data-driven simulation provides a significant, tangible estimate of the "knowledge deficit."

Collectively, the large gap between the solid red line (Iran's actual path) and the plausible scenarios modeled by the blue and green lines provides a robust, data-grounded estimate of the significant and lasting cost of the 1979 disruption.

\subsection{Comparing Iran to Two Empirical Counterfactuals}

Figure~15 extends the counterfactual analysis introduced in Figure~14 by incorporating two empirically grounded benchmarks: (i) a Synthetic Iran constructed via the Synthetic Control Method (SCM), and (ii) a scaled version of South Korea’s historical trajectory calibrated to Iran’s pre-1979 level. Together, these models allow us to quantify both a conservative and an ambitious estimate of Iran’s “knowledge deficit” by 2024.

\begin{figure}[htbp]
    \centering

    \begin{subfigure}[t]{0.8\textwidth}
        \centering
        \includegraphics[width=0.8\textwidth]{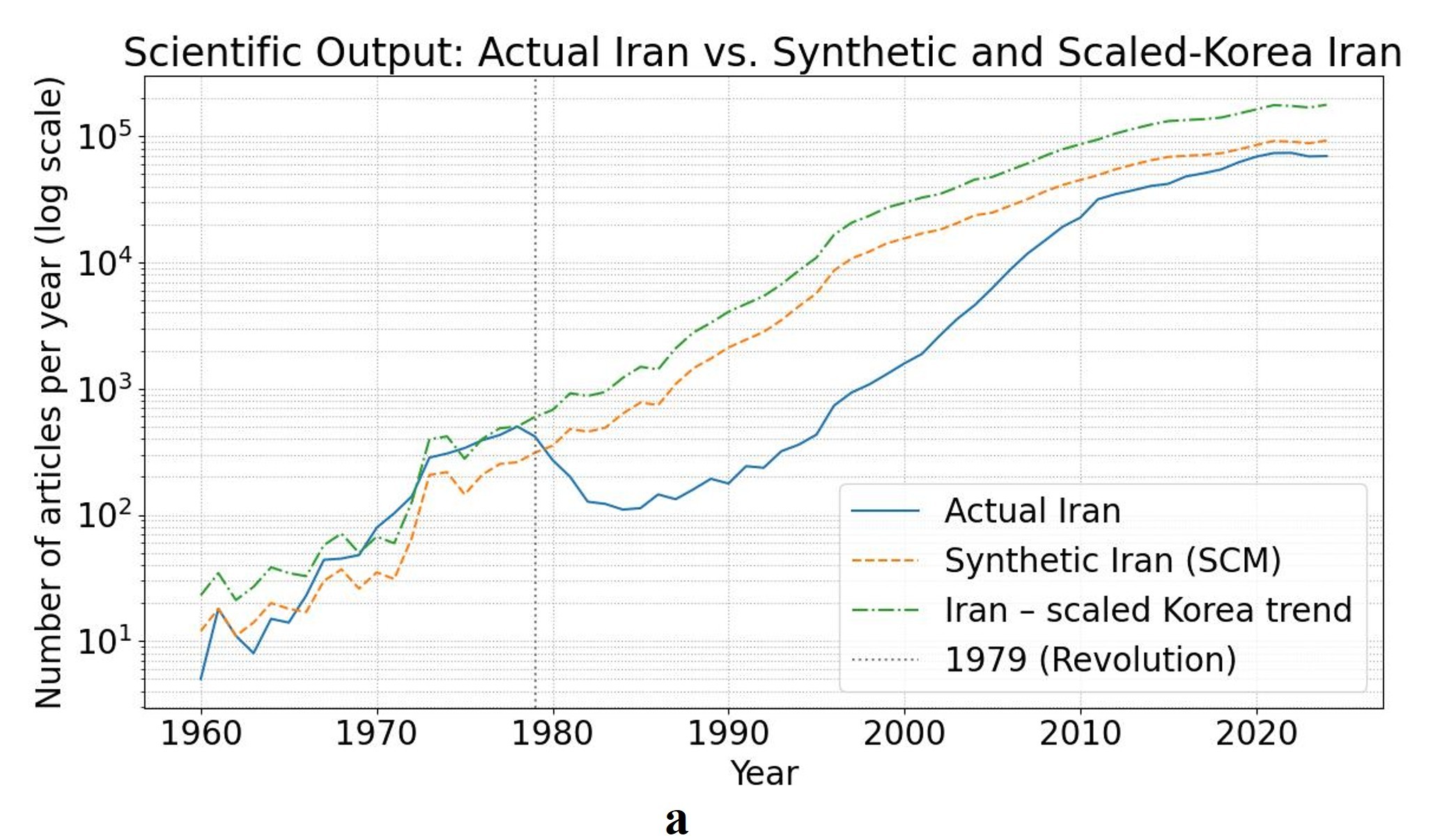}
        \caption{Scientific Output: Actual Iran vs. Synthetic and Scaled-Korea Iran.}
        \label{fig:15a}
    \end{subfigure}

    \vspace{0.5cm}

    \begin{subfigure}[t]{0.8\textwidth}
        \centering
        \includegraphics[width=0.8\textwidth]{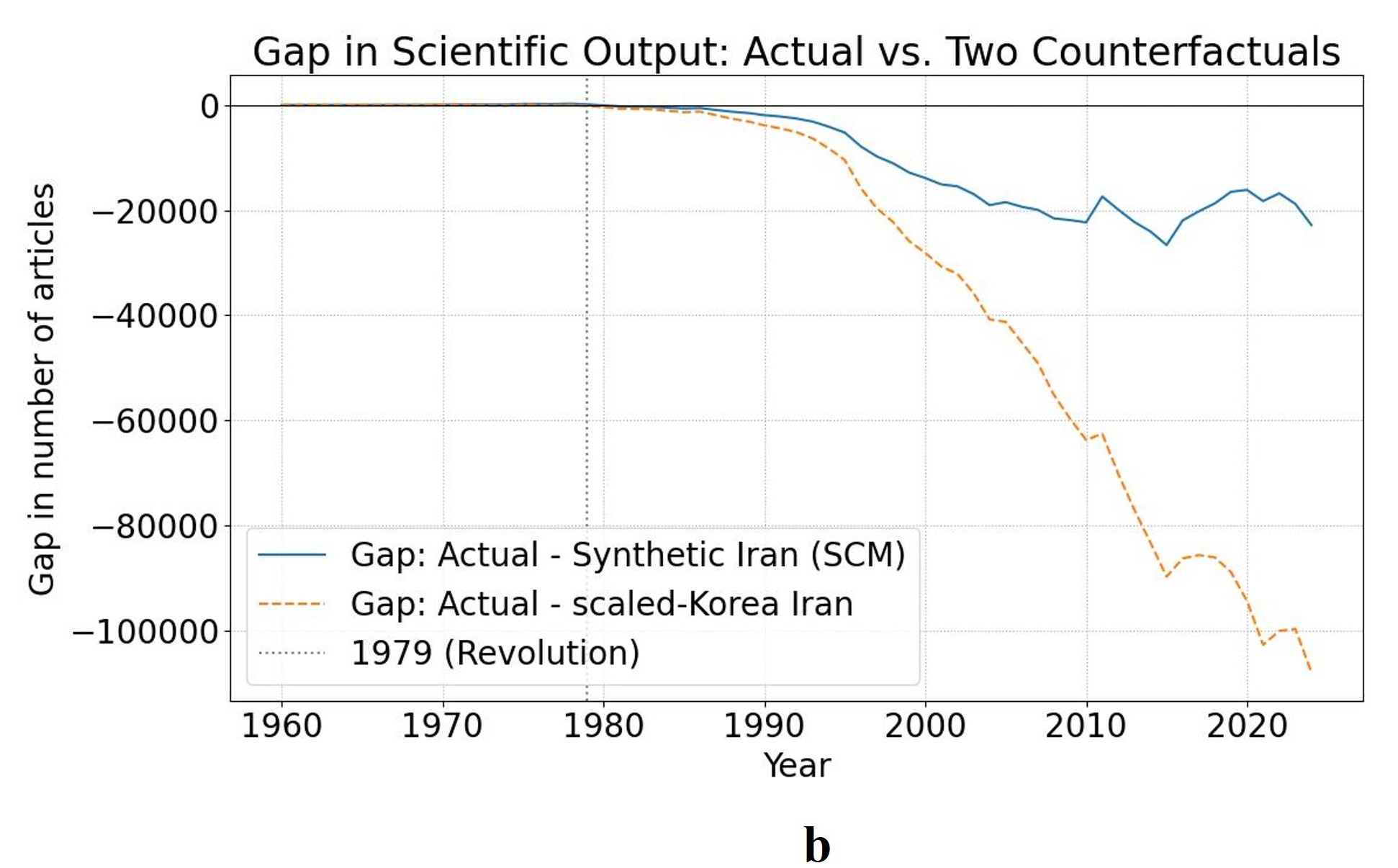}
        \caption{Gap in scientific output relative to Synthetic Iran (SCM) and scaled-Korea counterfactuals.}
        \label{fig:15b}
    \end{subfigure}

    \vspace{0.05cm}

    \begin{subfigure}[t]{0.8\textwidth}
        \centering
        \includegraphics[width=0.8\textwidth]{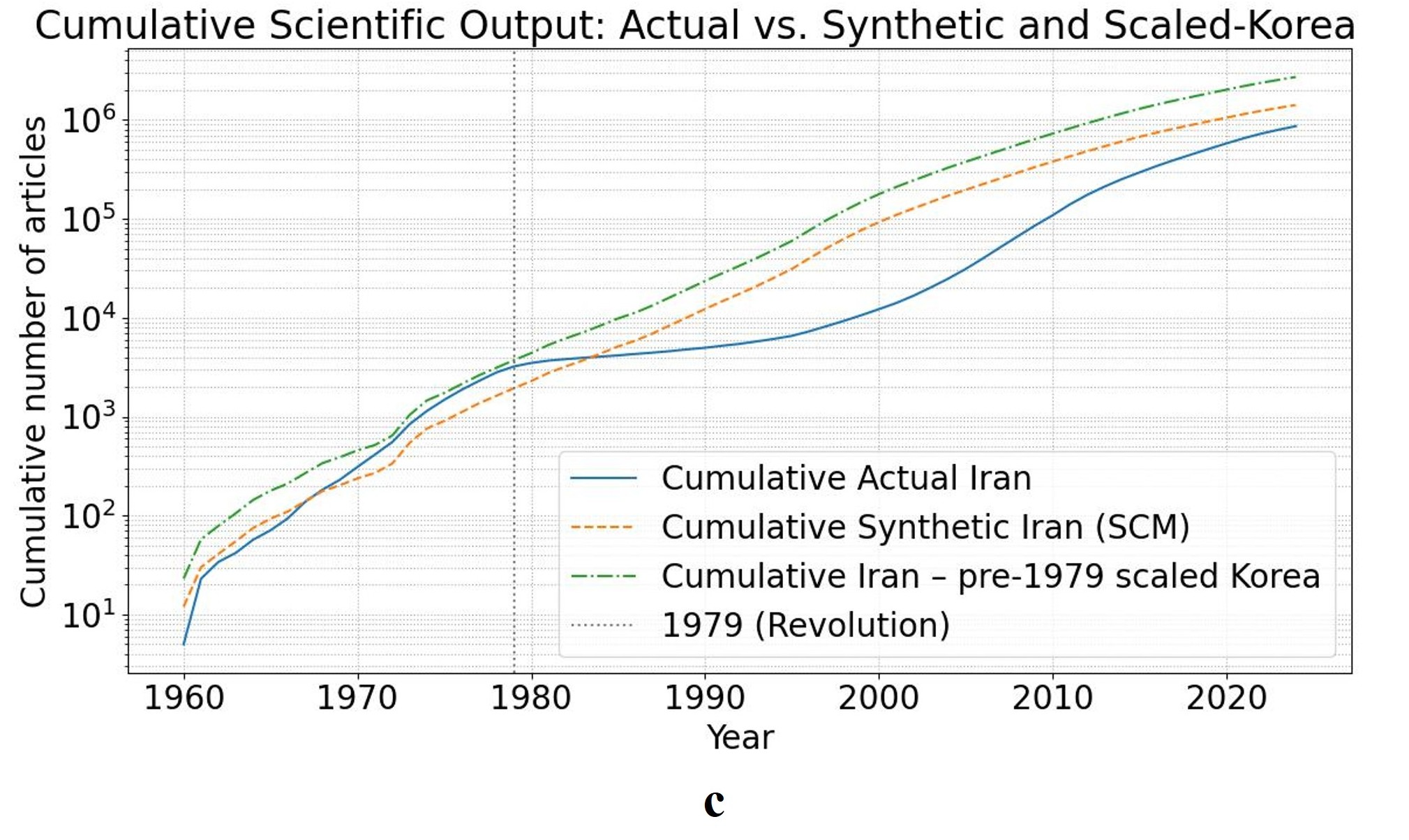}
        \caption{Cumulative scientific output under actual and counterfactual trajectories.}
        \label{fig:15c}
    \end{subfigure}

    \caption{Comparative evaluation of Iran’s scientific output under two data-driven counterfactuals.}
    \label{fig:15}
\end{figure}

It is important to clarify how the modelling strategy in Figure~14 differs from the 
synthetic–control framework adopted in Figure~15. Figure~14 presents three 
data–driven but explicitly model–based counterfactual trajectories. All three 
paths rely on real empirical information, but they remain stylized projections designed to illustrate 
a range of plausible, data–anchored futures.

By contrast, Figure~15 implements a fully non–parametric, comparison–based 
approach using the Synthetic Control Method (SCM). Rather than assuming a 
functional form or imposing a specific growth structure, SCM constructs a 
counterfactual Iran directly from the weighted combination of real countries 
that most closely match Iran’s pre–1979 economic and educational profile. 
Whereas the models in Figure~14 illustrate how Iran might have evolved under 
different empirically informed assumptions, the SCM in Figure~15 provides a 
data–driven comparative benchmark grounded solely in observed pre–treatment 
similarities. Together, these complementary approaches allow us to triangulate the magnitude of Iran’s post–1979 divergence using both model–based and 
comparison–based counterfactual evidence.

The SCM is a methodology widely used in economics and political science for estimating the causal impact of major historical shocks~\autocite{abadie2010synthetic}. The core idea of SCM is to approximate the “treated unit”---here, Iran after 1979---with a weighted combination of untreated comparison countries whose pre–intervention characteristics closely resemble those of the treated unit. Unlike traditional regression methods, SCM does not impose a functional form; instead, it builds the closest possible replica of Iran’s pre–1979 trajectory using convex weights that are optimized to minimize the pre–treatment prediction error.

In this study, the SCM model is constructed using annual publication counts from Scopus as the dependent variable and four structural predictors from the World Bank database: GDP per capita (constant prices), tertiary–education enrollment, government expenditure, and trade openness. The model is trained strictly on the pre–1979 period, ensuring that the synthetic version of Iran is calibrated exclusively to the historical conditions that existed before the Revolution. 

The optimization procedure assigns a weight of 1.0 to South Korea as the
sole donor country. This outcome is consistent with the data structure of
our predictor set: among all candidate countries, Korea is the only one
whose pre–1979 trajectory in per–capita GDP, tertiary enrollment, public
expenditure, and trade openness jointly reproduces Iran’s pre–revolution
profile with minimal prediction error. In other words, the SCM algorithm
selects Korea not because of any arbitrary modeling choice, but because
it is the unique country in the donor pool whose multidimensional
economic–education vector most closely approximates Iran’s historical
values. This makes Korea the statistically optimal donor unit for
constructing a credible counterfactual.

The resulting “Synthetic Iran” represents an empirical estimate of how Iran’s scientific output would likely have evolved in the absence of the 1979 disruption, assuming its pre–Revolution structural fundamentals had remained stable. This counterfactual serves as a conservative benchmark for evaluating the long–term impact of the Revolution and the prolonged interruption of higher–education institutions on Iran’s scientific productivity.

Panel (a) shows that prior to 1979, all three trajectories follow a similar rising path, reflecting the shared developmental momentum of Iran and the Asian Tigers. Immediately after the Revolution and the Cultural Revolution–era shutdown of universities, the observed curve collapses while both counterfactual trajectories continue rising smoothly. The SCM path---which uses only economic and educational predictors in the pre-1979 period—grows moderately and represents a lower-bound estimate of Iran’s potential. The scaled-Korea curve grows substantially faster, reflecting the historically high compounding rates achieved by South Korea during 1980\-/2020.

Panel (b) quantifies the annual “gap” between actual output and each counterfactual. Both gaps remain close to zero before 1979 but widen significantly afterward. By the early 2000s the deficit becomes structurally embedded, and by 2024 the gap relative to the SCM counterfactual reaches roughly 22,847 articles, while the gap relative to the scaled-Korea benchmark exceeds 107,889. This clearly demonstrates that Iran’s post-1979 recovery never closed the structural distance accumulated during the lost decade.

Panel (c) displays cumulative output---a metric that captures the compounding nature of scientific production capacity. Whereas Iran’s cumulative total reaches approximately 864,675 papers by 2024, the SCM counterfactual produces around 1.42 million, and the scaled-Korea model exceeds 2.71 million. The cumulative “knowledge deficit” therefore ranges from a conservative estimate of 551,129 lost publications (SCM) to a substantially larger deficit of 1.85 million (scaled-Korea). The large divergence in cumulative terms underscores that the primary long-term cost of the 1979 disruption was not solely the immediate collapse of output, but the permanent loss of compounding scientific momentum.

In sum, Figure~15 demonstrates that even under the most conservative assumptions, Iran’s scientific system today represents only a fraction of its plausible counterfactual potential. The SCM model anchors a lower-bound scenario based on structural predictors, while the scaled-Korea trajectory approximates a realistic upper-bound scenario consistent with the growth path of Iran’s closest developmental analogue. Both converge on the same conclusion: the Revolution-induced shock resulted in a generational loss of Iran’s scientific capacity.

\section{Conclusion}

This paper provides a comprehensive, data–driven reconstruction of Iran’s scientific trajectory over more than six decades and evaluates its long–term evolution through two independent counterfactual frameworks. The findings demonstrate that the disruption triggered by the 1979 Revolution produced a structural break whose effects extend beyond the immediate decline in scientific activity. Rather than representing a temporary interruption, the post–1979 decade reset the underlying growth rate of Iran’s scientific system and altered its long–run development path.

The counterfactual analyses clarify the magnitude of this divergence. The Synthetic Control Method, which selects South Korea as the single optimal pre–1979 match for Iran, indicates that Iran’s scientific output would have followed a substantially steeper growth trajectory in the absence of institutional upheaval. A second model, constructed by scaling Korea’s post–1979 trajectory to Iran’s 1979 baseline, yields an even more accelerated path. Both counterfactuals converge on the same conclusion: by 2024, Iran’s actual scientific output remains far below the level that would have been expected had its pre–1979 momentum continued uninterrupted.

The comparison between these reconstructed trajectories and Iran’s observed performance highlights a key insight: Iran’s scientific system successfully regained scale in the post–2000 period, but it did so along a shallower curve than its historical peers. The resulting “knowledge deficit” is therefore not merely the product of lost years, but the cumulative consequence of a lost growth rate. The analysis shows that once a nation’s scientific institutions lose continuity, the long–term effects compound—even in the presence of later recovery efforts.

In sum, the evidence demonstrates that Iran’s scientific development has been shaped not only by its capacity to produce publications but by the structural conditions that enable sustained, compounding growth. The counterfactual estimates presented in this study quantify the generational cost of the disruption that occurred in 1979 and underscore the central role of stability, openness, and institutional continuity in shaping national scientific trajectories. This suggests that reversing the long-term deficit identified here is contingent upon rebuilding these foundational conditions and re-establishing a growth environment capable of restoring the momentum lost more than four decades ago.

\section*{Data Availability}

The datasets generated and/or analyzed during the current study are available from the corresponding author on reasonable request.

\clearpage
\printbibliography

\clearpage

\end{document}